\documentclass[twocolumn,amsmath,aps,fleqn]{revtex4}
 
\usepackage{amssymb}
\usepackage{amsmath}
\usepackage{epsfig}
\usepackage{subfigure}
\usepackage{mathrsfs}
\usepackage{longtable}
\usepackage{enumerate}
\usepackage{multirow}
\usepackage[export]{adjustbox}

\newcommand{\be}{\begin{eqnarray}}
\newcommand{\ee}{\end{eqnarray}}
\newcommand{\Hu}{{\cal H}}

\newcommand{\p}{^{\prime}}
\newcommand{\pp}{^{\prime\prime}}
\newcommand{\dM}{\delta_\Delta}
\newcommand{\dx}{\delta\xi}
\newcommand{\dfs}{\delta_{f\sigma }}

 \allowdisplaybreaks[1]

\begin{document}

\title{A Fast Route to Modified Gravitational Growth}

\author{Tessa Baker}
\email{tessa.baker@astro.ox.ac.uk}
\affiliation{Astrophysics, University of Oxford, Denys Wilkinson Building, Keble Road, Oxford, OX1 3RH, UK}

\author{Pedro Ferreira}
\email{p.ferreira1@physics.ox.ac.uk}
\affiliation{Astrophysics, University of Oxford, Denys Wilkinson Building, Keble Road, Oxford, OX1 3RH, UK}

\author{Constantinos Skordis}
\email{skordis@nottingham.ac.uk}
\affiliation{School of Physics and Astronomy, University of Nottingham, University Park, Nottingham, NG7
2RD, UK}
\affiliation{Department of Physics, University of Cyprus, Nicosia 1678, Cyprus}

 
\begin{abstract}
The growth rate of the large-scale structure of the universe has been advocated as the observable {\it par excellence} for testing gravity on cosmological scales. By considering linear-order deviations from General Relativity, we show that corrections to the growth rate, $f$, can be expressed as an integral over a `source' term, weighted by a theory-independent `response kernel'. This leads to an efficient and accurate `plug-and-play' expression for generating growth rates in alternative gravity theories, bypassing lengthy theory-specific computations. We use this approach to explicitly show that $f$ is sensitive to a degenerate combination of modified expansion and modified clustering effects. Hence the growth rate, when used in isolation, is not a straightforward diagnostic of modified gravity.
\end{abstract}

\maketitle

\section{Introduction}
\label{section:intro}
\noindent 
The formation of large-scale cosmological structure is acutely sensitive to the nature of gravitational collapse. It has been argued that an accurate measure of the growth rate, $f(a)$, defined as
\begin{eqnarray}
f(a)&\equiv&\frac{d\ln \Delta_M(a)}{d\ln a} \label{fdef}  
\end{eqnarray}
where $\Delta_M(a)$ is the amplitude of the growing mode of matter density perturbations, can be used to constrain deviations from General Relativity (GR).

The method of choice for measuring $f(a)$ (or equivalently, $f(z)$) is through redshift space distortions \cite{DavisPeebles1983,Kaiser1987,Peacock2001,Guzzo2008}. 
The two-point correlation function of galaxies in redshift space is both anisotropic and scale-dependent, due to two competing effects: on small scales, the virialized motions of galaxies dominate over the Hubble flow, resulting in the elongation of the contours of the correlation function along the line of sight -- the {\it fingers of god} effect. On larger scales, gravitational infall leads to a squashing of the contours that is detectable on scales of \mbox{$10-30$ h$^{-1}$ Mpc}. There has been substantial progress in modelling this effect, both analytically \cite{Scoccimarro_2004, Taruya2013} and numerically \cite{Jenningsetal_2012, Jennings_2012}, and a number of systematic effects (non-linearity, the role of bias) have been studied. We will not discuss these difficulties here (see \cite{SimpsonPeacock2010} for further details); the focus of this paper is what we can learn from a measurement of $f(z)$ once it has been extracted from the data.

The current observational status of $f(z)$ is promising and intriguing. The surveys of \cite{Beutler_2012, Samushia_2012, Tojeiro_2012, Blake_2012, Torre_2013} have measured the growth rate from $z=0.2$ to $z=1.3$ on a range of scales, with errors of approximately 10-20\%. These measurements have provided decisive evidence for ruling out some extreme theories of modified gravity \cite{Raccanelli_2012}. We will show in this paper that other theories give rise to more subtle signatures that still lie within current experimental error bars; however, this situation should change with the next generation of galaxy surveys (see \textsection\ref{section:forecasts}).  

The growth rate is a particularly attractive observable from a theoretical point of view. For a start, we expect to measure it predominantly on scales where linear cosmological perturbation theory is valid. There is a battery of well-seasoned techniques associated with linear perturbation theory, and it is possible to adapt these for use with nearly all modified gravity theories \cite{BakeretalPPF}. Extending growth rate calculations to the mildly non-linear regime is possible \cite{Jennings_2012,Linder_Samsing_2013} but still in its infancy; furthermore, the reliance on theory-specific N-body simulations prevents one from making general statements about the effects of modified gravity on these scales.

A key advantage of $f(z)$ is that the range of scales probed is well inside the cosmological horizon, where the {\it quasistatic} approximation can be applied (see \cite{Silvestri2013} for a detailed discussion). This means that the dependence on extra degrees of freedom (an almost-inevitable feature of modified gravity) can be simplified and much of the time-dependence of the gravitational field can be discarded. As a result, the equations of motion for density perturbations and the growth rate are easy to work with.

In this paper we wish to explore the power of the growth rate as a probe of gravity. To do so, we first briefly introduce the quasistatic approximation in \S\ref{section:quasistatic}, and then use it in \textsection \ref{section:growth_rate} to derive a generalized evolution equation for $f(z)$. We show how, in the quasistatic regime, it depends on: a) modified gravitational clustering properties, and b) modifications to the background expansion history. In \textsection\ref{section:linear_response} we propose a simple and efficient method for linking the observable quantity $f\sigma_8(z)$ to functions parameterizing deviations from GR . In  \textsection\ref{section:forecasts} we discuss the trade-off between the degree of agnosticism about gravity a parameterization implements and the resulting constraints on it from growth rate data. A particularly convenient way of mapping specific gravity models onto the formalism of this paper is to use the Parameterized Post-Friedmann formalism \cite{BakeretalPPF}; we demonstrate this in \textsection\ref{section:PPF} by calculating a suite of examples. We discuss our findings in  \textsection\ref{section:discussion}. The busy reader may like to focus on \textsection\ref{section:linear_response} in particular.

\section{The Quasistatic Regime}
\label{section:quasistatic}
This paper focuses on gravitational collapse in the \textit{quasistatic regime}. This is defined as the range of length-scales which are sufficiently large enough for linear perturbation theory to be accurate, but still significantly less that the cosmological horizon length. This permits two approximations to be made:\begin{enumerate}
\item The consideration of significantly subhorizon scales implies that, when working in Fourier space, terms containing factors of ${\cal H}/k$ can be safely neglected.
\item On these spatial length-scales, the time derivatives of scalar perturbations are negligible relative to their spatial derivatives. Here `scalar perturbations' means both the usual gravitational potentials and any new perturbations not present in GR (eg. $\delta\phi$ for theories involving a new scalar field $\phi$).
\end{enumerate}
A careful discussion of these two features was presented in \cite{Silvestri2013}, which we will not repeat here.

Whilst at first our use of the quasistatic approximations may may seem to limit the application of our work, we emphasize that quasistatic scales dominate current and near-future galaxy redshift surveys. Testing modified gravity in the non-linear regime requires theory-specific N-body simulations. As mentioned in the introduction, these are only available for a limited handful of theories at present \cite{Schmidt2009, Zhao2011,Barreira2013,Li2013}.

An appealing feature of the quasistatic regime is that it allows many theories to be packaged in a simplified, generic form, as follows. Consider a gravity theory involving a single additional scalar degree of freedom, eg. Galileon gravity, $f\left(R\right)$ gravity or scalar-tensor theories. Perturbations of the scalar follow an equation of motion which, schematically, has the form:
\begin{align}
\delta\phi^{\prime\prime}+2\Hu\,\delta\phi^\prime+\left[k^2+a^2m(a)^2\right]\delta\phi&={\cal S}\left(a, \Phi, \Psi\right)
\label{jog}
\end{align}
where the effective mass of the scalar is set by its potential, $m(a)^2=\partial^2 V(\phi)/\partial\phi^2$, and the source term $\cal S$ depends upon the specifics of the theory in question \footnote{The schematic form of eq.(\ref{jog}) is probably not sufficiently general to cover all cases of interest. We are merely trying to convey the spirit of the derivation and the approximations involved, which can be applied more widely.}.

When the quasistatic approximations above are applied, eq.(\ref{jog}) reduces to an algebraic relation between perturbations of the scalar and the gravitational potentials:
\begin{align}
\delta\phi \approx\frac{{\cal S}\left(a, \Phi, \Psi\right)}{(k^2+a^2m(a)^2)}
\end{align}
This relation can then be used to eliminate $\delta\phi$ from the linearized gravitational field equations. Furthermore, terms in the linearized field equations containing $\delta\dot\phi$ can be dropped under point 2) above. One then finds that the Poisson equation and the `slip' relation between the two metric potentials can be written in the form:
\begin{align}
2\nabla^2\Phi&=\kappa a^2\,\mu(a,k)\,{\bar \rho}_M\Delta_M \label{Poisson}\\
\frac{\Phi}{\Psi}&=\gamma(a,k) \label{slip}
\end{align}
where we have defined two time- and scale-dependent functions,  $\mu(a,k)$ and $\gamma(a,k)$. Eqs.(\ref{Poisson}) and (\ref{slip}) can be thought of as a simple parameterization of modified gravity in the quasistatic regime: a theory corresponds to a particular choice of functional forms for $\mu(a,k)$ and $\gamma(a,k)$. A more detailed derivation of these relations can be found in \S\textrm{IVC} of \cite{BakeretalPPF}; for some theory-specific examples see \cite{deFelice,Bloomfield_HD}.

The results presented in this paper should apply to any theory for which the quasistatic reduction to eqs.(\ref{Poisson}) and (\ref{slip}) is valid. This covers any model with a single new scalar degree of freedom; note that this is \textit{not} restricted to only simple scalar field models. For example, the spin-0 perturbations of a new vector field or the St\"{u}ckelberg field used to restore Lorentz invariance to Ho\u{r}ava-Lifschtiz gravity both act as scalar degrees of freedom. We highlight that the entire broad family of Horndeski models is subject to our analysis  \cite{Horndeski, Deffayetetal_2011, deFelice}.

\section{The Linear Growth Rate in Modified Gravity} 
\label{section:growth_rate}
\noindent 
We will begin our calculations by clearly laying out how modifications to the gravitational field equations will affect the evolution of the growth rate of density perturbations, as defined in eq.(\ref{fdef}). Consider the pressureless matter component of the universe. Small inhomogeneities in the energy density, $\delta_M$, are defined through $\rho_M={\bar \rho}_M(1+\delta_M)$, where ${\bar \rho}_M$ is the mean energy density. 
In the conformal Newtonian gauge the evolution equations for the velocity potential $\theta$ (where the velocity perturbation is $v_i=\nabla_i\theta$) and the gauge-invariant density contrast \mbox{$\Delta=\delta+3\Hu (1+\omega)\theta$} are:
\begin{align}
\dot{\Delta}_M&=3(\dot\Phi+\Hu\Psi)-\theta_M\left[k^2+3(\Hu^2-\dot\Hu)\right]\label{Delta_evol}\\
\dot{\theta}_M&=-\Hu\theta_M+\Psi\label{theta_evol}
\end{align}
Eq.(\ref{Delta_evol}) is derived by combining eq.(\ref{theta_evol}) with the usual energy conservation equation for a pressureless fluid. We use dots to denote derivatives with respect to conformal time, and our conventions for the metric potentials are displayed in the perturbed line element: 
\begin{align}
ds^2=a^2(\eta)\left[-(1+2\Psi)d\eta^2+ (1-2\Phi)dx^idx_i \right]
\end{align}
In what follows, we will sometimes suppress the arguments of functions for ease of expression. We warn the reader that $\Omega_M$ should always be interpreted as a time-dependent quantity; we will use $\Omega_{M0}$ to indicate the fractional energy density in matter \textit{today}.

In the quasistatic regime (see \S\ref{section:quasistatic}) the $k^2$ term dominates eq.(\ref{Delta_evol}), so to a good approximation:
\begin{align}
\dot\Delta_M&\approx-k^2{\theta}_M
\end{align}
Differentiating this expression with respect to conformal time and combining it with eq.(\ref{theta_evol}) we obtain:
\begin{align}
\label{uio}
\ddot\Delta_M+\Hu\dot{\Delta}_M+k^2\Psi\approx 0
\end{align}
Combining eqs.(\ref{Poisson}), (\ref{slip}) and (\ref{uio}) and using the Friedmann equation to express \mbox{$\kappa a^2 {\bar \rho}_M =3{\cal H}^2\Omega_M$} 
we obtain:
\begin{eqnarray} 
\ddot\Delta_M+\Hu\dot{\Delta}_M-\frac{3}{2}{\cal H}^2\Omega_M\xi{\Delta}_M=0 \label{DeltaF}
\end {eqnarray}
where we have defined $\xi\equiv\mu/\gamma$. The quantity $\xi(a,k)$ will appear frequently throughout this paper; it is equal to $1$ in GR. For convenience we rewrite eq.(\ref{DeltaF}) using $x=\ln a$ as the independent variable:
\begin{eqnarray} 
\Delta^{''}_M+\left(1+\frac{\Hu^{'}}{\Hu}\right){\Delta}^{'}_M-\frac{3}{2}\Omega_M\xi{\Delta}_M=0 \label{DeltainF}
\end {eqnarray}
Primes denote derivatives with respect to $x$. It is helpful to convert this second-order equation for $\Delta_M$ into a first-order equation for the growth rate. Employing the usual definition of eq.(\ref{fdef}), we have \mbox{$f=\Delta_M^\prime/\Delta_M$}, and the consequential result \mbox{$\Delta_M^{\prime\prime}/\Delta_M=f^{\prime\prime}+f^2$}. In terms of $f$ eq.(\ref{DeltainF}) becomes:
\begin{align}
& f^\prime+q(x)\,f+f^2=\frac{3}{2}\Omega_M \xi \label{f_eq}\\
& \mathrm{where}\quad\quad q(x)=\frac{1}{2}\left[1-3\,\omega(x) (1-\Omega_M(x)\right] \label{qdef}
\end{align}
We have introduced a free function, $\omega(x)$, acting as an effective equation of state of the non-matter sector. The unperturbed expansion history of any dark energy or modified gravity theory can be written in the form of the usual GR Friedmann equation with a new fluid component, through a suitable choice of $\omega(x)$ \cite{KunzSapone2006,Skordis2008}. Also, note that $\xi$ can generally be a function of scale, so we must allow for a possible scale-dependence of the growth rate, \mbox{$f=f(x,k)$}; this is a common property of modified gravity theories which distinguish them from GR.  
 
Whilst the growth rate is of prime importance, in practice one actually measures the {\it density-weighted} or {\it observable growth rate}, $f\sigma_8(x,k)$, where $\sigma_8$ is the root-mean-square of mass fluctuations in spheres of radius $8$ h$^{-1}$Mpc \cite{PercivalWhite2009}. $\sigma_8$ evolves with the same growth factor $D(x,k)$ as the matter overdensity, ie.:
\begin{align}
\label{hjk}
\frac{\sigma_R (z)}{\sigma_{R}(z=0)}\simeq D(z,k=\frac{2\pi}{R})= \frac{\Delta_M (z,k=\frac{2\pi}{R})}{\Delta_{M}(0,k=\frac{2\pi}{R})} 
\end{align} 
where $R=8$h$^{-1}$Mpc. This will prove useful in \textsection\ref{subsection:response_fsig8}.


\section{The Linear Response Approach}
\label{section:linear_response}

We can assume that any viable theory of modified gravity must result in observables that match the $\Lambda$CDM+GR model to a high degree of accuracy. We then ask the question: what small deviations from $\Lambda$CDM+GR are still permissible within the error-bars of current and near-future experiments? We will answer this question by considering linear perturbations about the `background' solution of $\Lambda$CDM+GR, which corresponds to $\omega=-1,\,\mu=\gamma=\xi=1$. 

Our approach should not be confused with the standard cosmological perturbation theory of an FRW universe. We are {\it already} working within the context of \textit{spacetime} linear perturbation theory.
We are now going to perturb around $\Lambda$CDM+GR in \textit{model space} by assuming that the functions $\xi$ and $\omega$ source small deviations $\delta f$ from the growth rate predicted by the $\Lambda$CDM+GR model. We will see shortly that this an excellent approximation to the full solution of the non-linear eq.(\ref{f_eq}). 

For simplicity we will first investigate the impact of these small perturbations in model-space on the (unobservable) growth rate $f$, before extending our treatment to the observable $f\sigma_8$ in \textsection\ref{subsection:response_fsig8}. 

\subsection{The Response Function of the Growth Rate $f$}
\label{subsection:response_f}

We begin by decomposing $f$ into a zeroth-order part and a perturbation. As stated above, the zeroth-order solution is that of GR and hence is scale-independent, but the perturbation may not be:
\begin{align}
f(x,k)&=f_{GR}(x)+\delta f(x,k)
\end{align}
Likewise we perturb $\xi$, $\omega$ and $\Omega_M$ about their GR+$\Lambda$CDM values by writing 
\begin{align}
\xi&=1+\delta\xi(x,k)\nonumber \\
\omega&=-1+\beta(x) \nonumber \\
\Omega_M&=\Omega^{(0)}_M+\delta\Omega_M(x)
\end{align}
where $\Omega^{(0)}_M=\bar{\rho}_M/(\bar{\rho}_M+\bar{\rho}_\Lambda)$, and $\bar{\rho}_\Lambda$ is the energy density of the non-matter sector in the zeroth-order $\Lambda$CDM solution, ie. it evolves as a perfect fluid with equation of state $\omega=-1$. 
Substituting these expansions into eq.(\ref{f_eq}) and equating first-order parts (and continuing to suppress some arguments for clarity):
\begin{align}
\delta f^\prime+q_{GR}(x)\,\delta f+2f_{GR} \,\delta f&=\frac{3}{2}\Omega^{(0)}_M(x)\,\delta \xi (x,k) \nonumber \\
&+\frac{3}{2}(1+f_{GR})\,\delta \Omega_M \nonumber \\
&+\frac{3}{2}(1-\Omega^{(0)}_M)f_{GR}\,\beta
\label{df_eq}
\end{align}
where \mbox{$q_{GR}(x)=\frac{1}{2}\left[4-3\Omega_M(x)\right]$}. In Appendix \ref{app:u_deriv} we show that:
\begin{align}
\delta\Omega_M=3\Omega^{(0)}_M(1-\Omega^{(0)}_M)u(x)
\label{delta_om}
\end{align}
where $u(x)=\int_0^x \beta (x')dx'$ such that $u(0)=0$. Using this in eq.(\ref{df_eq}), we obtain:
\begin{align}
\delta f^\prime+q_{GR}(x)\,\delta f&+2f_{GR} \,\delta f=\frac{3}{2}\Omega_M^{(0)}(x)\,\delta \xi (x,k) \nonumber \\
&+\frac{3}{2}(1-\Omega_M^{(0)})\left[(1+f_{GR})3\Omega_M^{(0)} u+f_{GR}\beta\right]\label{df_eq11}
\end{align}
It is more convenient to work with the fractional deviation of the growth rate from the $\Lambda$CDM+GR prediction, which we define as:
\begin{align}
\eta (x,k) &=\frac{\delta f(x,k)}{f_{GR}(x)}=\frac{f(x,k)}{f_{GR}(x)}-1\label{etadef}
\end{align}
In terms of this new variable (and using the zeroth-order part of eq.(\ref{f_eq})), eq.(\ref{df_eq11}) becomes:
\begin{align}
 &\eta^\prime+\eta\left[f_{GR}+\frac{3}{2}\frac{\Omega_M^{(0)}}{f_{GR}}\right]=\frac{3}{2}\frac{\Omega_M^{(0)}}{f_{GR}}\delta S  \label{etaprime_eq}
 \end{align}
where:
\begin{align}
 &\delta S=\delta \xi+\frac{(1-\Omega_M^{(0)})}{\Omega_M^{(0)}}\left[3\,\Omega^{(0)}_M\,\left(1+f_{GR}\right) u+f_{GR}\beta\right] \label{deltaS}
\end{align}
This first-order equation can be solved using an integrating factor, leading to the expression:
\begin{align}
\eta(x,k)& = \frac{3}{2} \int^x_{-\infty} \,\frac{\Omega_M^{(0)}(\tilde{x})}{f_{GR}({\tilde x})}\;\delta S({\tilde x,k})\;K(x,\,{\tilde x}) d{\tilde x} \label{eta_int}
\end{align}
where:
\begin{align}
K(x,\,{\tilde x}) &= \mathrm{exp}\left\{-\int^x_{\tilde x} d{\bar x}\left[f_{GR}(\bar{x})+\frac{3}{2} \frac{\Omega^{(0)}_M({\bar x})}{f_{GR}({\bar x})}\right]\right\}\label{kernel}
\end{align}
We see that the solution for $\eta (x,k)$ takes the form of an integral over a `source' term $\delta S (\tilde{x},k)$ and a `kernel' $K(x, \tilde{x})$. Crucially, note that the kernel only depends on the $\Lambda$CDM+GR background, meaning that it is theory-independent and simple to calculate. Effectively, the kernel acts purely as a weighting function. \textit{The entire theory-dependence of the modified growth rate is encoded in the source term $\delta S(x,k)$} which at each moment in time (or $x$) is a degenerate combination of $\delta \xi$, $\beta$ and $u$. This makes clear how modifications to different parts of the gravitational field equations propagate through to affect the growth rate (see below).

As one might expect, we need to know the background solution we are expanding about (due to the factors of $\Omega_M^{(0)}$ and $f_{GR}$) in order to solve for the deviation $\eta(x,k)$. The background solution is found by solving eq.(\ref{f_eq}) with $\xi=1$. In general this must be done numerically, but as the computation is done in standard GR it is straightforward and rapid to calculate.

Let us interpret eq.(\ref{eta_int}) physically. It says that the fractional deviation from $f_{GR}$ is an integral from early times \mbox{$(x\rightarrow-\infty)$} up to the time of observation. One expects that an observer will be more sensitive to non-GR behaviour occurring at times recent to him/her than at high redshift. This sensitivity is encoded in the kernel $K(x, \tilde{x})$: it gives an exponential suppression factor depending on the interval between the time of the deviation from GR ($\tilde{x}$) and the time of observation ($x$). For deviations occurring at the time of observation there is no suppression, \mbox{$K(x, x)=1$}. Causality imposes that $K(x, \tilde{x})=0$ for $\tilde{x}>x$; clearly observables cannot be affected by deviations from GR that occur \textit{after} the time of observation.

We can also apply some physical interpretation to function $\delta S$ (eq.(\ref{deltaS})) that sources corrections to the GR growth rate. It contains three contributing factors:
\begin{itemize}
\item The first term, $\delta\xi$, can be interpreted as the modified clustering properties stemming from the modified Poisson equation (eq.(\ref{Poisson}));
\item The second term, $u(x)=\int_0^x\beta(x^\prime)\,dx^\prime$, arises from the modified expansion \textit{history};
\item The third term, $\beta(x)$, describes the \textit{instantaneous} modified expansion rate, i.e. at the time that $\delta S$ is being evaluated.
\end{itemize}
It is important to note that the background expansion rate of the universe contributes significantly to the modified growth rate, ie. $f(z)$ is not solely a probe of linear perturbations. The degeneracy between $\beta$ and $\dx$ was also highlighted in \cite{SimpsonPeacock2010}; the authors further considered the Alcock-Paczynski-related issues that a non-$\Lambda$CDM background would pose for the extraction of $f\sigma_8(z)$ from a galaxy power spectrum. The analysis of this paper has instead focussed on how to best use the growth rate signal once we have it in hand. 

\begin{figure*}[t]
\begin{center}
\subfigure{\label{fig:gauss}\includegraphics[scale=0.42]{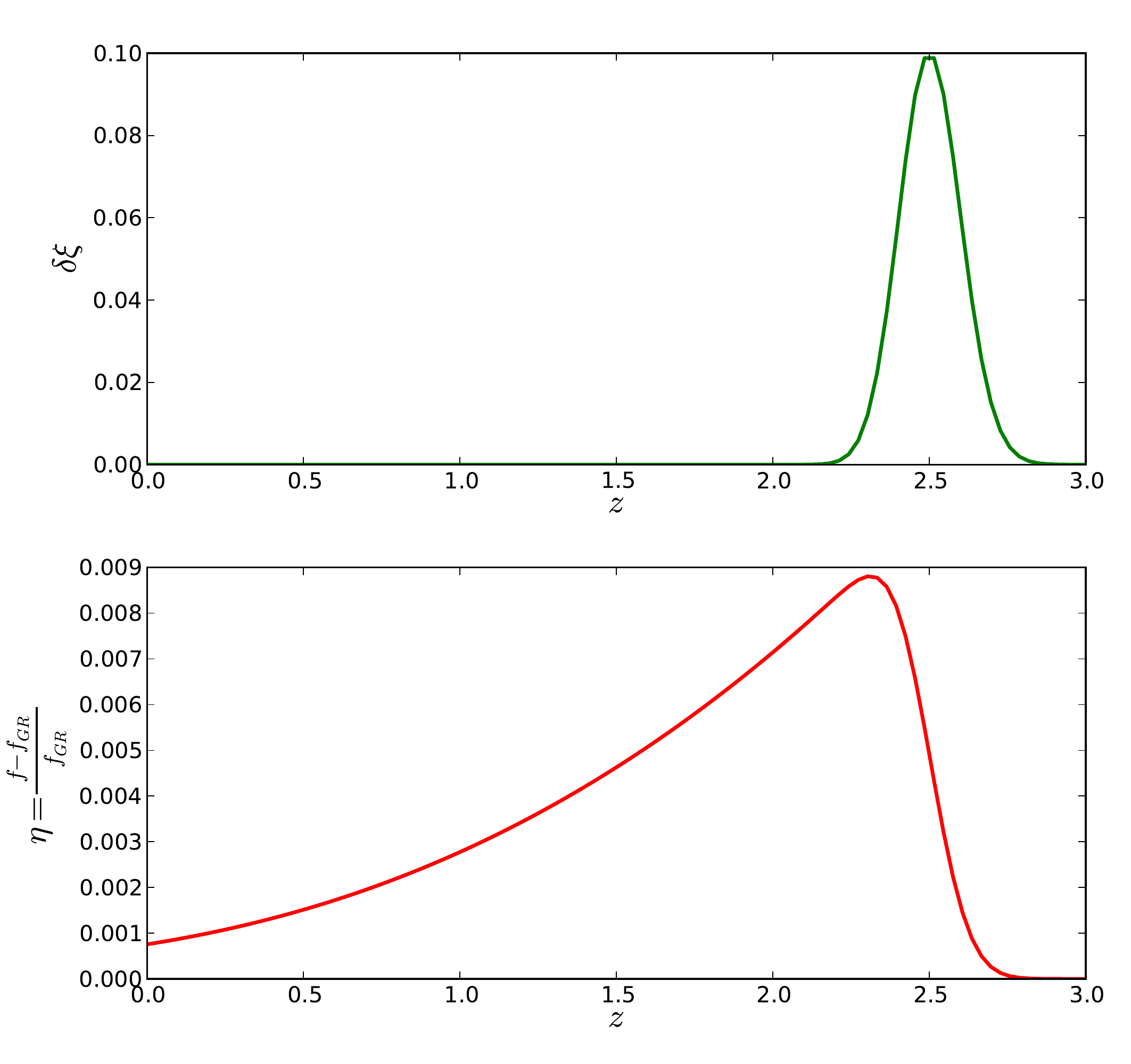}}%
\subfigure{\label{fig:step}\includegraphics[scale=0.42]{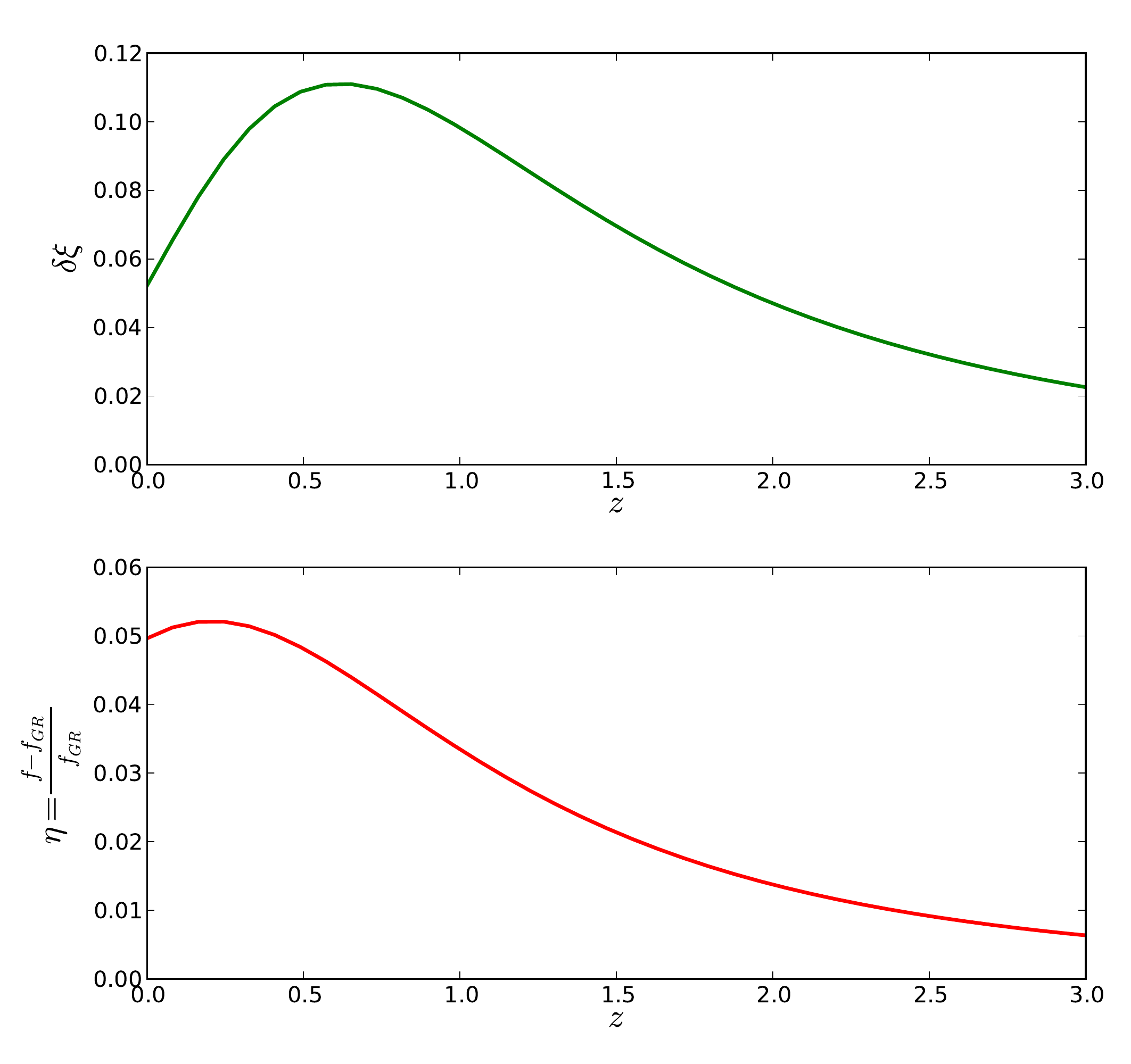}}%
\vspace{-1cm}
\end{center}
\caption{Examples of how the growth rate is affected by different source terms in eq.(\ref{f_eq}), where \mbox{$\delta\xi=\xi-1$}. Effects on the growth rate are expressed as a percentage deviation from the GR prediction, ie. \mbox{$\eta=f/f_{GR}-1$}. For the right-hand panel $\dx=1+0.75\,\left(1-\Omega^{(0)}_M\right)-1.5\,\left(1-\Omega^{(0)}_M\right)^2+0.75\,\left(1-\Omega^{(0)}_M\right)^3.$}
\label{fig:response_figs1}
\end{figure*}

We will put eq.(\ref{eta_int}) to use by considering some toy examples. These will illustrate the response of the growth rate to arbitrary deviations from GR; they are not intended to represent any particular theory of modified gravity. Let us first consider a simple case where the background expansion precisely matches that of a cosmological constant (i.e. \mbox{$\beta=0$}), and only the behaviour of perturbations is modified. 
Fig.~\ref{fig:response_figs1} shows the fractional deviation of the growth rate (defined in eq.(\ref{etadef})) triggered by two forms of modifications to matter clustering: a Gaussian and a cubic-order Taylor series in \mbox{$[1-\Omega^{(0)}_{M}(x)]$}, where $\Omega^{(0)}_M(x)$ evolves as predicted by the $\Lambda$CDM+GR model. In each case we have assumed a scale-independent $\delta\xi$, but one could construct a more complicated function of $k$ and consider Fig.~\ref{fig:response_figs1} as snapshots at a given scale. 

We see from the left panel of Fig.~\ref{fig:response_figs1} that after a transient $\delta \xi$ source the growth rate gradually returns to its GR value, decaying approximately as \mbox{$a^{-\frac{5}{2}}$} for a reasonably narrow Gaussian (the index can be inferred by considering eq.(\ref{eta_int}) during a matter-dominated epoch). The rate of return to GR is slightly suppressed at late times when the background expansion starts to accelerate, which acts to `freeze in' perturbations. 

Sustained modifications to GR such as those considered in the the right-hand panel of Fig.~\ref{fig:response_figs1} lead to growing deviations from $f_{GR}$, and hence will generally be more tightly constrained by data. For example, the cubic polynomial shown results in a $\sim6\%$ effect on the growth rate at \mbox{$z=0$}, substantially larger than the sub-percent deviations shown in the left-hand panel. 

Note that there is a time lag between changes in $\delta \xi$ and the response of the growth rate. For example, in the right-hand panel of Fig.~\ref{fig:response_figs1} the non-GR source begins to die away after $z\sim0.5$, but $\eta(z)$ has insufficient time to follow suit. One could imagine generalizations of this situation, in which GR is the correct description of our universe today, but the effects of past non-GR behaviour still persist for a limited time (such late-time changes in the dynamics of the dark sector were explored in \cite{Amin_2012}).

One may justifiably ask what kind of error is introduced by approximating eq.(\ref{f_eq}) as a linear equation. In fact the error is extremely small for the situations we are considering here. For an example with a $\Lambda$CDM background (such as the one shown in the left-hand panel of Fig.~\ref{fig:response_figs1}), the full (non-linearized) evolution equation for \mbox{$\delta f(x)=f(x)-f_{GR}(x)$} given by eq.(\ref{f_eq}) has a solution with the same form as eq.(\ref{eta_int}), but with a modified kernel:
\begin{align}
K^{Full}(x,\,{\tilde x}) &= \mathrm{exp}\left\{-\int^x_{\tilde x} d{\bar x}\left[f(\bar{x})+\frac{3}{2} \frac{\Omega_M({\bar x})}{f_{GR}({\bar x})}\right]\right\}
\end{align}
The first term of the integrand is now the modified growth rate instead of $f_{GR}$. That is to say:
\begin{align}
K(\tilde{x},x)&=K^{Full}(\tilde{x},x)\,\mathrm{exp}\left[-\int_{\tilde{x}}^x f_{GR}(\bar{x})-f(\bar{x})\,d\bar{x}\right]\nonumber\\
&=K^{Full}(\tilde{x},x)\,\mathrm{exp}\left[\int_{\tilde{x}}^x \delta f(\bar{x})\,d\bar{x}\right]
\end{align}
For the small deviations from $\Lambda$CDM+GR the exponential factor above is of order unity, so the linearized kernel is a very good approximation to the real kernel for $\eta$. For the examples shown in Fig.~\ref{fig:response_figs1} the error on $\eta (z)$ introduced by linearizing the growth rate equation is $\leq 1\%$. We emphasize that here we are \textit{not} talking about a $1\%$ error on the growth rate $f(z)$; we are talking about a $1\%$ error on $\delta f/f_{GR}$, a quantity that is itself a small percentage of the growth rate.

\subsection{The Response Function of the Observable Growth Rate $f\sigma_8$}
\label{subsection:response_fsig8}

Extending the linear response analysis of the previous subsection to the observable growth rate,  $f\sigma_8$, is straightforward. We will continue to focus on scale-independent modifications. 
Using eq.(\ref{hjk}), the fractional deviation of $f\sigma_8$ from its value in GR is given by:
\begin{align}
\delta_{f\sigma}&=\frac{\delta[f\,\sigma_8](z)}{f\sigma_8|_{GR}(z)}\nonumber\\
&=\frac{\delta f(z)}{f(z)_{GR}}+\frac{\delta\sigma_8(z)}{\sigma_8(z)|_{GR}}\nonumber\\
&=\eta(z,k)+\frac{\delta\Delta_M(z,k)}{\Delta_M(z)|_{GR}}
\label{frac_fs8}
\end{align}
where the first equality defines $\delta_{f\sigma}$, and $\delta\Delta_M$ is the deviation of the gauge-invariant matter density perturbation from its corresponding value in the $\Lambda$CDM+GR scenario. We have already calculated the first term in the last line above, so we now tackle the second. For convenience we define a new symbol for this:
\begin{align}
\delta_\Delta (z,k)&=\frac{\delta\Delta_M(z,k)}{\Delta_M(z)|_{GR}}
\end{align}
 By perturbing eq.(\ref{DeltainF}) about the $\Lambda$CDM+GR model we obtain: 
\begin{align}
\dM^{\prime\prime}+\dM^\prime\left(1+\frac{ \Hu^\prime}{\Hu}+2f_{GR}\right)=\frac{3}{2} \Omega_M^{(0)}\,\delta S (x) \label{DeltaMx}
\end{align}
where $\delta S$ is again given by eq.(\ref{deltaS}). 

Eq.(\ref{DeltaMx}) can be solved for $\delta_\Delta^\prime$ using an integrating factor, then integrated once more to obtain $\delta_\Delta$. Reversing the order of the integrations allows us to write the solution in the form `source $\times$ kernel', as we did for $\eta$ in the previous subsection:
\begin{align}
\dM(x,k)=\frac{3}{2}\int_{-\infty}^x\,\Omega_M^{(0)}(\tilde{x})\,\delta S(\tilde{x},k)\,I(x,\tilde{x})\,d\tilde{x}
\label{rty}
\end{align}
where, as before, the kernel $I(x,\tilde{x})$ is a function of the zeroth-order $\Lambda$CDM cosmology only:
\begin{align}
\label{Iintx}
I(x,\tilde{x})&=\int_{\tilde{x}}^x dy \,\mathrm{exp}\left[-\int^y_{\tilde{x}}\,d\bar{x}\left(2-\frac{3}{2}\Omega_M^{(0)}(\bar{x})+2f_{GR}(\bar{x})\right)\right]
\end{align}
and we have used the Friedmann equation for the GR background en route.

\begin{figure*}[t]
\begin{center}
\hspace{-5mm}
\subfigure{\label{fig:cubic}\includegraphics[scale=0.52]{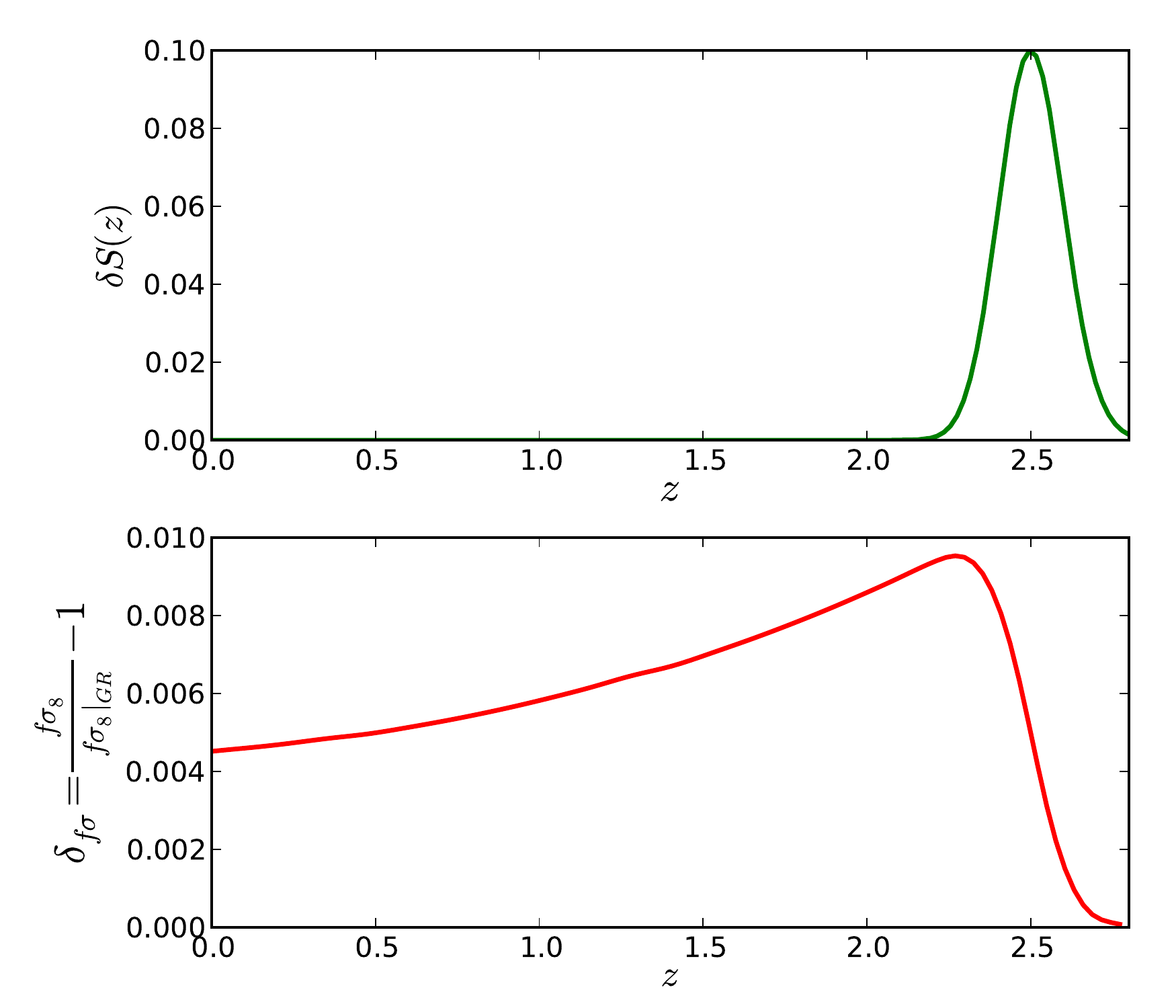}}
\subfigure{\label{fig:bkgd}\includegraphics[scale=0.52]{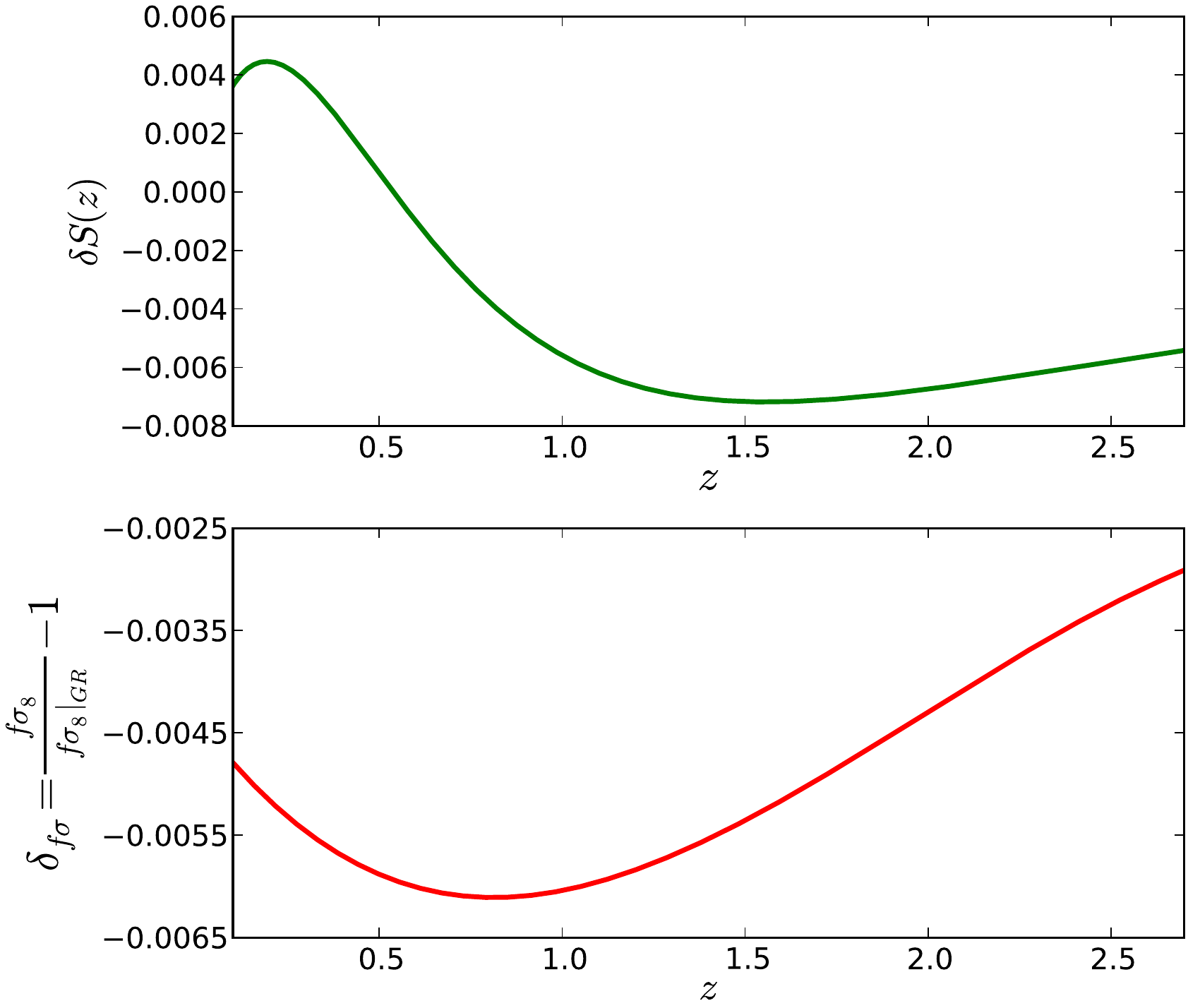}}%
\end{center}
\vspace{-1cm}
\caption{Lower panels show the fractional deviation of the density-weighted growth rate from its $\Lambda$CDM+GR model caused by the source functions in the upper panels, see eq.(\ref{fs8_int}). The left panel shows the same Gaussian considered in the left panel of Fig.~\ref{fig:response_figs1}. The right panel shows the effect of an evolving background of the CPL variety, that is, \mbox{$\omega(a)=\omega_0+\omega_a (1-a)$}. We have fixed \mbox{$\omega_0=-1,\,\omega_a=0.05$}, which is equivalent to \mbox{$\beta(x)=0.05(1-e^x)$}. Note that $\delta S\rightarrow 0$ at high redshifts (not shown) since \mbox{$\Omega^{(0)}_{M}(z)\rightarrow 1$} there (see eq.(\ref{deltaS})).} 
\label{fig:fsig8_figs}
\end{figure*}

Finally, using eq.(\ref{frac_fs8}) and the results of \textsection\ref{subsection:response_f}, the fractional deviation of $f\,\sigma_8$ from its $\Lambda$CDM+GR value can be expressed in a Green's function-like form:
\begin{align}
\label{fs8_int}
\frac{\delta[f\sigma_8(x,k)]}{f\sigma_8(x)|_{GR}}&=\int_{-\infty}^x\,\delta S(\tilde{x},k)\,G(x,\tilde{x})\,d\tilde{x}
\end{align}
where the kernel $G(x,\tilde{x})$ is:
\begin{align}
\label{Gkernel}
G(x,\tilde{x})&=\frac{3}{2}\Omega_M(\tilde{x})\left[\frac{K(x,\tilde{x})}{f_{GR}(\tilde{x})}+I(x,\tilde{x})\right]
\end{align}
and the factors $K(x,\tilde{x})$ and $I(x,\tilde{x})$ are given by eqs.(\ref{kernel}) and (\ref{Iintx}). Fig.~\ref{fig:fsig8_figs} shows uses of this formula. The left-hand panel shows the same case considered in the left panel of Fig.~\ref{fig:response_figs1}, where $\dx$ has a Gaussian form. Whilst $\eta(z)$ declined to zero, $\delta_{f\sigma}(z)$ settles to a constant. The difference in behaviour arises from the second term of eq.(\ref{frac_fs8}), as follows: during the time the source $\delta S$ is `switched on' the growth of density perturbations is either enhanced or suppressed relative to the $\Lambda$CDM+GR case. When the source switches off density perturbations return to growing at the GR \textit{rate}, but their absolute value has now been shifted from that of a pure $\Lambda$CDM+GR universe. This shift is the constant term seen in Fig.~\ref{fig:fsig8_figs}.

The right-hand panel of Fig.~\ref{fig:fsig8_figs} shows the effect of allowing the background effective equation of state to evolve too, ie. \mbox{$\beta(x)\neq0$}. In this plot we have considered a CPL-like equation of state, that is:
\begin{align}
\omega(a)=\omega_0+\omega_a (1-a)
\end{align}
but left the clustering properties of GR unaffected (ie. \mbox{$\dx=0$}). Clearly a realistic analysis would need to allow both \mbox{$\dx\neq 0$} and \mbox{$\beta\neq 0$} simultaneously. We have treated the two contributions separately here to compare them: note that whilst the upper plots of Fig.~\ref{fig:fsig8_figs} have roughly the same peak amplitude, the right-hand panel shows a much larger impact on $f\sigma_8(z)$. This is because $f\sigma_8$ is sensitive to a time-integrated effect (see eq.(\ref{fs8_int})), and an evolving background equation of state constitutes a more sustained source ($\delta S$) than the transient Gaussian shown in the left-hand panel.

Eqs.(\ref{fs8_int}) and (\ref{Gkernel}) are a key result of this paper, so let us summarize what has been achieved. Accepting that the $\Lambda$CDM+GR model is an excellent description of the universe at leading order, we have found a general way to calculate the impact that modifications to the General Relativistic field equations have on the observable growth rate of structure. All the modifications are encapsulated in a single function $\delta S(x,k)$, which can be matched to a specific gravity theory or constrained in a model-independent, phenomenological manner. 

We will see in \textsection\ref{section:PPF} that for fully-specified gravity models $\delta S(x,k)$ depends only on \textit{background-level} quantities of that theory, so there is no need to perform the full perturbation analysis. This represents a significant decrease in the mathematical workload. Similarly, our formalism bypasses the need to write a separate growth rate numerical code for every gravity theory of interest; a simple background solver is enough (this usually amounts to solving a few uncomplicated ODEs).

On a practical note, we re-iterate that the kernel $G(x,\tilde{x})$ is a function of the zeroth-order $\Lambda$CDM cosmology only, and hence is relatively simple to compute. It only needs to calculated once and stored (as a function of $x$ and $\tilde{x}$) to allow rapid calculation with different source functions. Furthermore, for the examples considered in this paper, we have found our method is remarkably accurate; the error on $\dfs$ incurred using our linearized treatment (instead of the exact calculation) is of order $2\%$ for the example shown in the left-hand panel of Fig.~\ref{fig:fsig8_figs}, and of order $10\%$ for the right-hand panel \footnote{The error is larger for sustained forms of $\delta S$ than for transient ones due to the longer integration time. Small nonlinear terms, neglected in our treatment, have a larger cumulative effect in the sustained case.}. This is equivalent to a small fraction of a percentage error on $f\sigma_8 (z)$ itself, well within the accuracies forecast for next-generation galaxy surveys.


\section{Parameterization vs. Constraints -- the Trade-Off}
\label{section:forecasts}

We now proceed to show how the formalism of the previous sections be connected to galaxy redshift surveys. Let us assume that we have measurements of $\dfs$ (defined in eq.(\ref{frac_fs8})) from a survey in $N$ redshift bins, with centers $x_i$ (recall \mbox{$x=\mathrm{ln}a$}) and widths $w_i$, where \mbox{$i=1,\ldots,N$}. The first step is to discretize eq.(\ref{fs8_int}):
 \begin{align}
\dfs^i &= \sum_{j=1}^i \left[\int_{x_j-\frac{w_j}{2}}^{x_j+\frac{w_j}{2}}\,G(\tilde{x},x)\,d\tilde{x}\times \delta S_j\right]=\sum_{j=1}^iG_{ij}\,\delta S_j 
\label{eta_discrete}
\end{align}
$\dfs^i$ and $\delta S_i$ are vectors containing the mean values of $\dfs(x)$ and $\delta S(x)$ in each redshift bin; for the present we will assume negligibly weak scale-dependence over the range of interest (the linear regime). $G_{ij}$ is a triangular matrix due to the causality requirement discussed in \textsection\ref{section:linear_response}.

Next we must consider how to choose the quantities we wish to constrain. We will see below, as is often the case with parameterized methods, that one must strike a balance between the generality of the parameterization and the size of the error-bars obtained on the parameters/functions involved. Inputting more information (via, for example, constraint equations or specifying the time/scale-dependence of free functions) will result in a parameterization that is more tightly constrained but less widely applicable.

We will investigate three degrees of parameterization:
\begin{itemize}
\item {\bf Full agnosticism} -- one simply constrains the source function $\delta S_i$ in each redshift bin, where $i=1,\ldots, N$.
\item {\bf Compromise} -- one uses eq.(\ref{deltaS}) for $\delta S$ and supplies a functional form for $\dx (x)$ and $\beta (x)$. The functional form used can be an approximation over the redshift range relevant to the observations. Effectively, one is inputting some prejudices about how we expect modifications to GR to evolve with time, without being specific about the origin of those modifications. In this case one constrains the set of $M$ parameters used in specifying the form of $\delta S (x)$. We will denote these collectively as $\{\dx_a\}$, where $a=1,\dots, M$.
\item {\bf Model-specific} -- one uses the field equations of a particular gravity theory to express $\dx (x)$ and $\beta (x)$ in terms of the Lagrangian parameters of that theory. We will denote the Lagrangian parameters to be constrained by $\{\lambda_y\}$, where $y=1,\ldots, R$. 
\end{itemize}

We assume a multivariate Gaussian likelihood with mean zero for the deviation of the density-weighted growth rate measurements from $\Lambda$CDM, that is:
\begin{align}
\ln{\cal L}&=-\ln\left[(2\pi)^{N/2}\,\prod_{i=1}^N\sigma^i_{f\sigma_8}\right]-\frac{1}{2}\chi^2\\
\mathrm{where}\quad \chi^2&=\sum_{i=1}^N\frac{\left(f\sigma_{8,\,\Lambda CDM}^i-f\sigma_{8, \, th}^i\right)^2}{(\sigma_{f\sigma_8}^i)^2}
\end{align}
$f\sigma_{8,\,th}^i$ is the density-weighted growth rate in bin $i$ predicted by a modified gravity theory, and $(\sigma^i_{f\sigma_8})^2$ is the experimentally-determined variance on $f\sigma_8$ in that bin.

\begin{figure}[tb!]
\begin{center}
\includegraphics[scale=0.53]{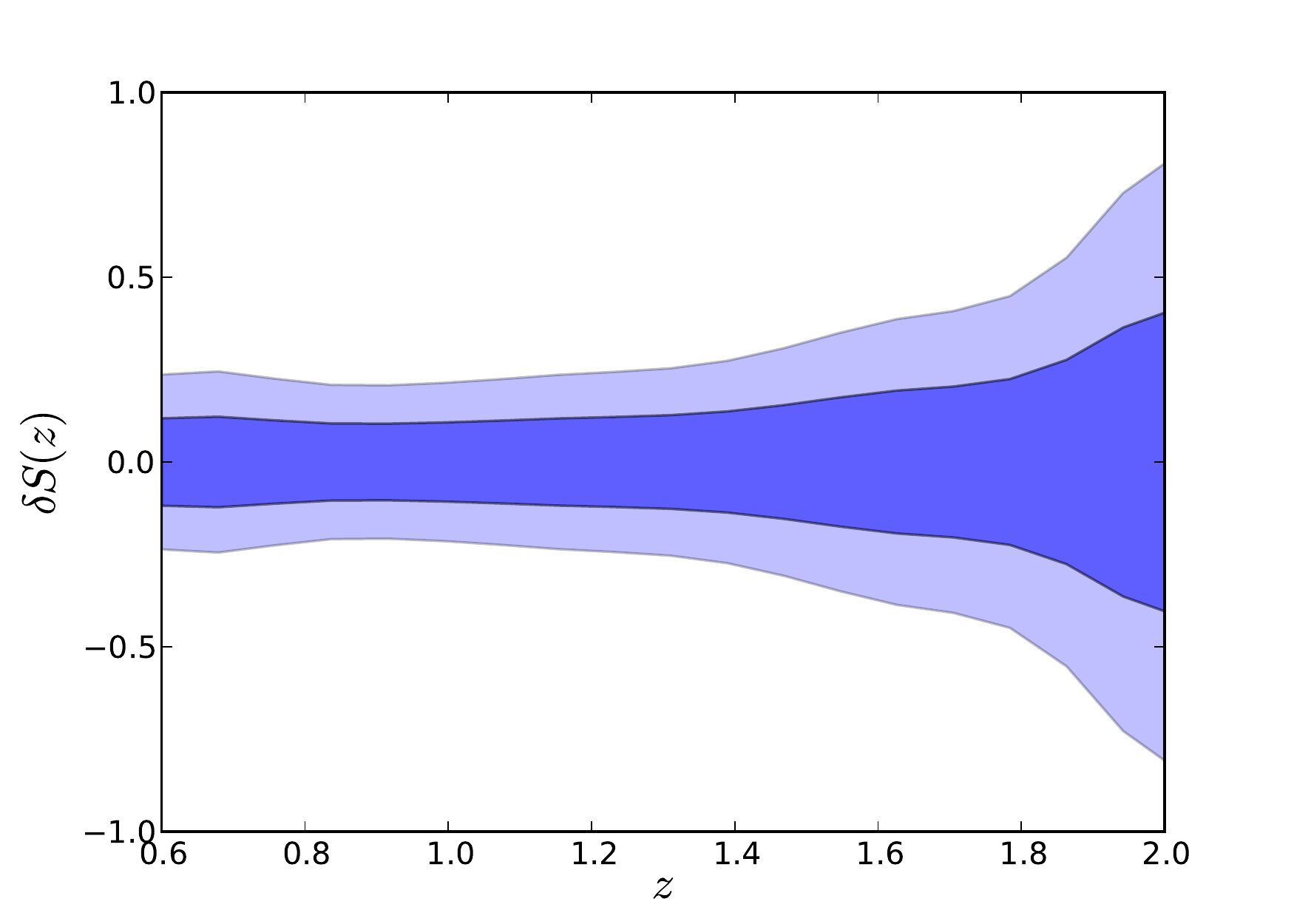}
\caption{Forecast constraints on $\delta S$ for a typical next-generation galaxy survey, where $\delta S$ sources deviations from the GR growth rate (see eq.(\ref{deltaS})). 
The contours shown represent $1\,\sigma$ and $2\,\sigma$ uncertainties. }
 \label{xi_forecast}
 \end{center}
 \end{figure}

First consider constraining a `fully agnostic' parameterization. Using the standard formalism of Fisher matrices, the inverse covariance matrix for the $\delta S_i$ is given by:
\begin{align}
 ({ C^{\delta S}})^{-1}_{ij}&={ F^{\delta S}_{ij}}=\langle\partial_{\delta S_i}\partial_{\delta S_j}\chi^2\rangle\\
&=\Bigg\langle\sum_{k=1}^N\frac{1}{\left(\sigma^k_{f\sigma_8}\right)^2}\,\frac{\partial(f_{\sigma_8,\,th}^k)}{\partial\,\delta S_i}\frac{\partial(f_{\sigma_8,\,th}^k)}{\partial\,\delta S_j}\Bigg\rangle
\label{rkf}
\end{align}
To obtain the derivatives above we make use of eq.(\ref{eta_discrete}):
\begin{align}
f\sigma_{8,\,th}^k&=f\sigma_{8,\,\Lambda CDM}^k+\delta f\sigma_8^k\\
&=f\sigma_{8,\,\Lambda CDM}^k\left(1+\sum_{j=1}^N G_{kj}\,\delta S_j\right)\\
\Rightarrow\quad \frac{\partial(f_{\sigma_8,\,th}^k)}{\partial\,\delta S_i}&=f\sigma_{8,\,\Lambda CDM}^k\,G_{ki}
\label{swn}
\end{align}
Using eq.(\ref{swn}) in eq.(\ref{rkf}) and defining:
\begin{align}
\tilde{\sigma}^k_{f\sigma_8}=\sigma^k_{f\sigma_8}/f\sigma_{8,\,\Lambda CDM}
\end{align}
we obtain:
\begin{align}
{ F^{\delta S}}_{ij}&=\sum_{k=1}^N\frac{1}{(\tilde\sigma^k)^2}\,G^T_{ik}\,G_{kj}
\label{mwb}
\end{align}
For this simplified analysis we will neglect correlations between the redshift bins of our survey. Then, by defining the diagonal covariance matrix ${\Sigma}_{ij}=1/(\tilde{\sigma}^k)^2\,\delta_{ij}$, we can rewrite eq.(\ref{rkf}) in matrix form:
\begin{align}
\left(\mathbf C^{\delta S}\right)^{-1}&={\mathbf F^{\delta S}}={\mathbf G^T\,\mathbf\Sigma\,\mathbf G}
\end{align} 
Inverting $\mathbf F^{\dx}$ yields the covariance matrix of interest. Fig.~\ref{xi_forecast} shows a representative example for the constraints on $\delta S$ as a function of redshift obtained using a next-generation RSD survey. The results are surprisingly uninformative -- an input $\sim 1\%$ error on measurements of $f\sigma_8$ has resulted in a $\sim 10\%$ error on $\delta S$ (at 1$\sigma$ confidence).

\begin{figure}[tb!]
\hspace{-1cm}
\includegraphics[scale=0.48, left]{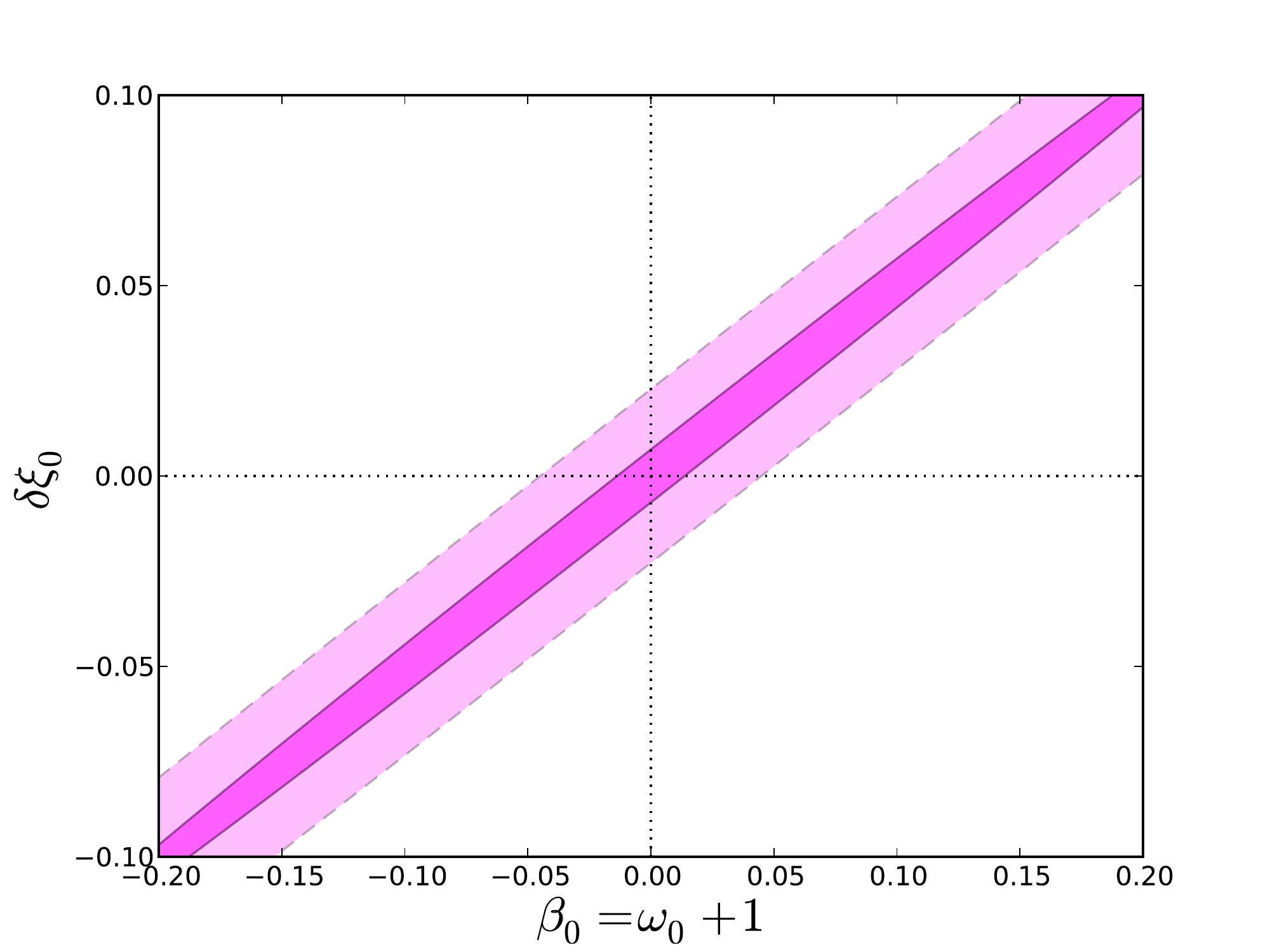}
\caption{Constraints on the simplest example of a `compromise' parameterization (described in the text) using a next generation RSD survey. $\beta_0$ is a (constant) deviation of the equation of state from $-1$. $\delta\xi_0$ encapsulates the novel clustering properties of a non-GR gravity theory. The contours shown represent $1\,\sigma$ and $2\,\sigma$ uncertainties. }
 \label{level2}
 \end{figure}

Now let us investigate what happens when we impose more restrictions on the parameterization. Let us consider the simplest possible example of a `compromise' parameterization, in which we take the functions $\dx (x)$ and $\beta (x)$ to be approximately constant over the redshift range of interest. Using eq.(\ref{deltaS}) we then have:
\begin{align}
\label{dS_simple}
\delta S&=\delta \xi_0 +\alpha(x)\beta_0 
\end{align}
where $\delta \xi_0$ and $\beta_0$ are constants and: 
\begin{align}
\label{alpha_simple}
\alpha(x)&=\frac{\left(1-\Omega_M^{(0)}\right)}{\Omega_M^{(0)}}\left[3\,\Omega_M^{(0)}x\, (1+f_{GR})+f_{GR}\right] 
\end{align}
The covariance matrix for the parameters $\dx_a=\{\dx_0,\beta_0\}$ is given by an expression analogous to eq.(\ref{rkf}), but with the derivatives now being taken with respect to the parameters $\{\dx_a\}$. The chain rule allows us to rewrite this as:
\begin{align}
& ({ C^{\dx}})^{-1}_{ab}={ F^{\dx}_{ab}}\\
&=\Bigg\langle\sum_{k=1}^N \sum_{i=1}^N \sum_{j=1}^N\frac{1}{\left(\sigma^k_{f\sigma_8}\right)^2}\,\frac{\partial \delta S_i}{\partial\,\dx_a}\frac{\partial(f_{\sigma_8,\,th}^k)}{\partial\,\delta S_i}\frac{\partial(f_{\sigma_8,\,th}^k)}{\partial\,\delta S_j} \frac{\partial\delta S_j}{\partial\,\dx_b}\Bigg\rangle
\label{ksn}
\end{align}
Repeating steps similar to the fully agnostic case, we reach the matrix expression:
\begin{align}
\left(\mathbf C^{\dx}\right)^{-1}&={\mathbf F^{\dx}}={\mathbf P^T \mathbf G^T\,\mathbf\Sigma\,\mathbf G\,\mathbf P}
\label{led}
\end{align} 
where $\mathbf P$ is the $N\times M$ Jacobian matrix \mbox{$ \displaystyle P_{ia} = \left[ \frac{\partial\delta S_i}{\partial\,\dx_a}\right]$}. For the example of eqs.(\ref{dS_simple}) and (\ref{alpha_simple}) this is simply 
\begin{align}
\mathbf P = \begin{pmatrix} 1 & \alpha_1 \\
1 & \alpha_2 \\
\vdots & \vdots \\
1 & \alpha_N
\end{pmatrix}
\end{align}
where $\alpha_i$ are discretized values of $\alpha (x)$ (from eq.(\ref{alpha_simple})), evaluated at the midpoint of each redshift bin. Fig.~\ref{level2} shows the constraints obtained on the model of eq.(\ref{dS_simple}) using the same representative next-generation RSD survey as Fig.~\ref{xi_forecast}. The plot makes plain the benefit of combining growth rate measurements with probes of the background expansion rate. If $\beta_0$ is left free, we have a degeneracy between positive values of $\beta_0$ -- which make the effective dark energy sector more important at earlier times, suppressing structure formation -- and positive values of $\delta\xi_0$ (which enhance structure growth). However, if we can pin \mbox{$\omega=-1$} to $1\%$ accuracy using other data, we can achieve similar $\sim 1\%$ constraints on deviations from GR.

Finally we consider a model-specific analysis, where the situation becomes a little more subtle. The quantities $\{\delta S_i\}$ and $\{\dx_a\}$ were related by a linear transformation, which preserves the (assumed) Gaussian nature of the joint probability distribution for either set of parameters. This enabled us to move straightforwardly from eq.(\ref{rkf}) to eq.(\ref{ksn}) via the chain rule. However, in general the parameters $\{\dx_a\}$ will not be linearly related to the model-specific Lagrangian parameters $\{\lambda_y\}$. This means that we cannot assume a Gaussian probability distribution for the parameter set $\{\lambda_y\}$, and it would be risky to continue applying a Fisher matrix analysis.

Instead we will make use of the normalized probability distribution that we have already obtained for the `compromise' parameterization. We re-express the ({non-constant}) $\dx(x)$ and $\beta(x)$ in terms of the Lagrangian parameters of a particular theory (one way to do this is via the PPF formalism, see \textsection\ref{section:PPF}). The elements of the covariance matrix for $\{\lambda_y\}$ can then be calculated directly, ie.
\begin{align}
\mathbf C^{\lambda}=&\int\hdots\int\,d\lambda_y\,d\lambda_z\hdots d\lambda_R\times\nonumber\\
&\frac{1}{n} \,\lambda_y\lambda_z\,\mathrm{exp}\left[-\frac{1}{2}\vec{u}^T(\mathbf \lambda)\,\mathbf F^{\dx} \,\vec{u} (\mathbf \lambda)\right]
\label{level3C}
\end{align}
where $n$ is a normalization factor. $\vec{u}(\lambda)$ is a column vector of length $2N$; it holds the expressions for of $\dx(x)$ and $\beta(x)$ in terms of the parameters $\{\lambda_y\},$ evaluated in each redshift bin.

Table~\ref{level3} gives some examples of model-specific constraints obtained using eq.(\ref{level3C}) and the same survey specifications as Figs.~\ref{xi_forecast} and \ref{level2}. The relevant expressions for these theories are given in \textsection\ref{section:PPF}.  It is difficult to exactly quantify the results of Table~\ref{level3} as percentages since these parameters are all zero in the GR limit (compare to $\delta S,\,\dx$ and $\beta$, which we know to be perturbations about quantities that are of order unity in GR). 

However, one generally obtains tighter constraints than those of the agnostic or compromise parameterizations, because model-specific expressions severely restrict how $f\sigma_8(z)$ is allowed to evolve from one redshift bin to the next. We now see the aforementioned trade-off between generality and constraining power at work. As we added more information into the analysis, moving from fully agnostic $\rightarrow$ compromise $\rightarrow$ model-specific cases, our $2\,\sigma$ constraints decreased. Hence an advantage of the formalism presented in this paper is its flexibility: the user can choose where they would like to position themselves on the sliding scale of generality versus constraining power.
\begin{center}
\begin{table}
\begin{tabular}{| c | c | c | c |}
\hline
\bf{Theory} & \bf{Parameter} & \bf{Fiducial value} & \;\;\;$\quad 2 \sigma \quad $\;\;\; \\
\hline
Brans-Dicke & $1/\omega_{BD}$ & 0.0 & $4.19\times 10^{-4}$ \\ \hline
\multirow{3}{*}{Einstein-Aether} & $c_1$ & 0.0 & 0.551  \\
 & $c_3$ & 0.0 & 4.312 \\
& $\alpha$ & 0.0 & 0.606 \\ \hline
DGP & $1/\left(\tilde{r}_cH_0\right)$ & 0.0 & 0.004 \\ \hline
\end{tabular}
\caption{Constraints obtained on model-specific Lagrangian parameters for three example theories, using the procedure described in \textsection\ref{section:forecasts}.}
\label{level3}
\end{table}
\end{center}

As a final comment, our formalism makes it clear that parameters will always be degenerate within a single redshift slice of a survey. Let us define the matrix $\mathbf A=\mathbf P^T \mathbf G^T$, so that eq.(\ref{led}) can be written as \mbox{$\mathbf F^{\dx}=\mathbf A\,\mathbf \Sigma\,\mathbf A^T$}. If we only have one redshift bin, this becomes:
\begin{align}
\label{degen}
\mathbf F^{\dx}&=\frac{1}{\tilde{\sigma}_{f\sigma_8}^2}\vec{A}\,\vec{A}^{\,T}
\end{align}
where $\vec{A}$ denotes a column of $\mathbf A$ and $\vec{A}\,\vec{A}^{\,T}$ is an outer product. It can be shown that the matrix formed by taking the outer product of a vector with its transpose is always singular (the reader may like to briefly consider any two-dimensional example). The singularity of $\bf F^{\dx}$ in this case implies that one of its eigenvectors has the eigenvalue zero. This is equivalent to a direction of complete degeneracy in the parameter space of $\{\dx_a\}$.

This situation is rectified by combining different redshift bins. Eq.(\ref{degen}) then becomes a sum, which prevents the singularity of the matrix $\mathbf F^{\dx}$:
\begin{align}
\label{degen}
\mathbf F^{\dx}&=\sum_{k=1}^N\frac{1}{\left(\tilde{\sigma}^k_{f\sigma_8}\right)^2}\vec{A_k}\,\vec{A_k}^{\,T}
\end{align}
where $\vec{A}_k$ denotes the $k$th column of $\mathbf A$. Eq.(\ref{degen}) is equivalent to eq.(\ref{led}).

\section{Deriving $\delta\xi$ and $\beta$ from the Parameterized Post-Friedmann Formalism}
\label{section:PPF}

In this paper we have aimed to keep our treatment of modified gravitational growth as theory-independent as possible. 
So far we have required only that the quasistatic approximation be valid for some range of scales, and made use of the widely-applicable quasistatic equations (\ref{Poisson}) and (\ref{slip}). Nevertheless, one may often be interested in testing a particular gravity theory. In this section we describe how fully-specified theories map onto our general formalism. We will do this first of all by using the Parameterized Post-Friedmann framework (PPF) of \cite{BakeretalPPF} (not to be confused with a different work of the same name by other authors \cite{Hu:2007fw}). However, we stress that it is \textit{not} obligatory to use PPF to apply the earlier results of this paper.

\subsection{The Quasistatic Limit of PPF}
PPF, inspired by the well-established Parameterized Post-Newtonian formalism (PPN) \cite{ThorneWill1971, Will1971, Will2006}, is a framework for model-independent tests of deviations from the field equations of GR using cosmological data. It describes the mathematically possible extensions of the linearized field equations (modulo some very mild restrictions, see \cite{BakeretalPPF}) in terms of a set of redshift-dependent functions. This set of functions acts as a cosmological analogy to the set of ten PPN parameters: different theories of modified gravity correspond to different specifications of them. And just like the PPN parameters, they can be constrained by calculating observable quantities and comparing to data.

The quantity $\xi=\mu/\gamma$ that appears in eq.(\ref{f_eq}) can be written in terms of the PPF coefficient functions. For the present we will consider theories which: 
\begin{enumerate}[a)]
\item contain no higher than second-order time derivatives in their equations of motion (a generic, but not absolute, stability criterion \cite{Woodard});\newline
\item contain one new non-GR degree of freedom, which we denote by $\chi$ (this could be a spin-0 perturbation of a new field, for example).
\end{enumerate}

The quasistatic form of the PPF field equations in the conformal Newtonian gauge is:
 \begin{align}
-a^2\delta G^0_0&=\kappa a^2 G\,\rho_M\delta_M+A_0 k^2\Phi+\alpha_0k^2\hat\chi\label{FE1}\\ 
-a^2\delta G^0_i&=\nabla_i\left[\kappa a^2 G\,\rho_M (1+\omega_M)\theta_M+B_0 k\Phi+\beta_0 k\hat\chi\right]\label{FE2}\\
a^2\delta G^i_i&=3\,\kappa a^2 G\,\rho_M\Pi_M+C_0 k^2\Phi+\gamma_0 k^2\hat\chi\label{FE3}\\
a^2\delta \tilde{G}^i_j&=\kappa a^2 G\,\rho_M (1+\omega_M)\Sigma_M+ D_0\Phi+\epsilon_0\hat\chi \label{FE4}
\end{align}
where $\delta\tilde{G}^i_j=\delta G^i_j-\frac{1}{3}\delta^i_j\,\delta G_k^k$ and $D_{ij}=\vec{\nabla}_i\vec{\nabla}_j-1/3 \delta_{ij}\vec{\nabla}^2$. Linear pressure perturbations are denoted by \mbox{$\delta P_M=\rho_M \Pi_M$} and $\Sigma_M$ is an anisotropic stress perturbation, which we will neglect hereafter. 
The hat over $\hat\chi$ indicates that it is a gauge-invariant combination of perturbations constructed using the algorithm of \cite{BakeretalPPF}, ie. it contains both $\chi$ and metric perturbations.

The alphabetic and Greek coefficients in eqs.(\ref{FE1})-(\ref{FE4}) are \textit{not} constants; they are functions of time and scale, but we have suppressed those arguments here for clarity. These are the PPF coefficients that one maps a theory of gravity onto. In fact, the scale-dependence of these functions is fixed \cite{BakeretalPPF,Silvestri2013,Amendola2013}, so they can be considered purely as functions of time.

For many theories the equation of motion (hereafter e.o.m.) of $\chi$ corresponds to a conservation equation \footnote{More specifically, it is the time-like component of the equation enforcing the divergenceless nature of tensor additions to the field equations of GR. Theories with explicit matter-scalar coupling introduce some subtlety here.}. This includes scalar-tensor gravity, quintessence, Einstein-Aether theory, Ho\u{r}ava-Lifschitz gravity and theories which fall into the Horndeski class \cite{Horndeski,Deffayetetal_2011,Gao_Steer_2011,deFelice}. When expressed in terms of the PPF coefficients, the quasistatic limit of the e.o.m. for $\chi$ is:
\begin{align}
\label{chi_eom}
&\hat\chi \left[\dot{\alpha}_0+k\beta_0\right]+\Phi\left[\dot{A}_0+k B_0\right]=0
\end{align}
Combining eq.(\ref{chi_eom}) with eqs.(\ref{FE1}) and (\ref{FE4}), the connection between the quasistatic $\{\mu,\gamma\}$ parameterization and the PPF functions is:
\begin{align}
\mu (z,k)&\approx \left\{1+\frac{A_0}{2} -\frac{\alpha_0}{2}\left(\frac{\dot{A}_0+k B_0}{\dot{\alpha}_0+k\beta_0}\right)\right\}^{-1}\label{muQS}\\
\gamma (z,k)&\approx \left\{1-D_0 +\epsilon_0\left(\frac{\dot{A}_0+k B_0}{\dot{\alpha}_0+k\beta_0}\right)\right\}^{-1}
\label{gammaQS}
\end{align}
The authors of \cite{Silvestri2013} recently derived a result relating the two quasistatic functions $\{\mu,\gamma\}$ for the Horndeski class of theories. When converted into our notation, their result is equivalent to the statement that the numerator of $\mu$ must be equal to $1$, which is manifest in eq.(\ref{muQS}).

Below we give some examples of the modified clustering and background functions, $\delta\xi$ and $\beta$, that were utilized in \textsection\ref{section:growth_rate} and \textsection\ref{section:linear_response}. Although eqs.(\ref{muQS}) and (\ref{gammaQS}) were derived for theories with only one non-GR degree of freedom, more complicated theories can still be mapped onto specifications of $\dx$ and $\beta$ under the quasistatic approximation. However, they need to be treated on a case-by-case basis rather than via eqs.(\ref{muQS}) and (\ref{gammaQS}) -- see the example of DGP below.

We should also highlight an interesting subtlety here with regards to the popular family of $f\left(R\right)$ models. It is often assumed that any results pertaining to scalar-tensor theories automatically incorporate $f\left(R\right)$ gravity, since a conformal mapping exists between the two classes of theories. Whilst this is true at the action level, the perturbed e.o.m.s for the new degree of freedom derived from their actions are \textit{not} equivalent. The e.o.m. of scalar-tensor theory is a conservation equation of the kind described above~\footnote{That is, the equation of motion of the scalar field is equivalent to the time-like component of the conservation law for the effective stress-energy tensor of the scalar. In the language of \cite{BakeretalPPF} these are `Type 1' theories.}; the e.o.m. for the `scalaron', $\delta f_R$ originates from the trace of the $f\left(R\right)$ gravitational field equations. Hence eqs.(\ref{muQS}) and (\ref{gammaQS}) do not apply to $f\left(R\right)$ gravity. Nevertheless we can still take the quasistatic limit of the theory, see below.\newline

\subsection{Examples}
\label{sub:QS_examples}
Here we present some examples for the clustering function $\dx$ and the background modification $\beta$. The relevant actions and references can be found in \cite{BakeretalPPF}. Note that all the expressions below are functions of the modified cosmological background only, making them relatively simple to evaluate. 

We also highlight that $\dx$ is scale-independent in all the cases presented here, except possibly scalar-tensor theory, where it depends on the choice of potential $V(\phi)$. We will not impose the requirement that the modifications to the field equations are the sole cause of acceleration. For example, we treat the `normal branch' of DGP gravity, which requires a cosmological constant in addition to the brane-based modifications.\\

\noindent\textit{\textbf{Scalar-Tensor Theory}}
\begin{align}
\label{xiST}
&\dx_{ST}(a,k)=-1+\left[\phi+\frac{\dot\phi}{\Hu}-\frac{Y}{Z}\right]\times\nonumber\\
 &\quad\quad\quad\quad\quad\left[\phi+\frac{1}{2}\left(1-\frac{a^2}{k^2}V\p(\phi)\right)\left(\frac{\dot\phi}{\Hu}-\frac{Y}{Z}\right)\right]^{-1}\\
&\mathrm{where}\quad Y=\frac{a^2}{k^2}V\p(\phi)\left(\frac{\ddot\phi}{\Hu}-\frac{\dot\Hu}{\Hu}\dot\phi\right)+\frac{\dot\phi^2}{\Hu}\frac{a^2}{k^2}V\pp(\phi)\nonumber\\
&\quad\;\;\quad\quad\quad+\dot\phi\left(\frac{\omega(\phi)\,\dot\phi}{\Hu \phi}-3\right)\\
&\quad\quad\quad Z=\frac{a^2}{k^2}\left(V\pp(\phi)\dot\phi+2\Hu V\p(\phi)\right)+\omega\frac{\dot\phi}{\phi}-\Hu\\
&\beta_{ST}(a)=\frac{2(\Hu^2-\dot\Hu)(1-\phi)+\omega(\phi)\frac{\dot\phi^2}{\phi}+\ddot\phi-2\Hu\dot\phi}{3\Hu^2(2-\Omega_{M0}^{(0)}-\phi)+\frac{1}{2}\omega(\phi)\frac{\dot\phi^2}{\phi}-3\Hu\dot\phi+a^2\,V(\phi)}
\label{betaST}
\end{align}
The scale-dependence of $\dx$ in this case has arisen because we have been careful not to make any assumptions about the form of the potential $V(\phi)$. It is likely that once a form is chosen for $V(\phi)$ further terms can be dropped due to the quasistatic approximation.
\vspace{0.5cm}

\noindent\textit{\textbf{Brans-Dicke Theory}}\newline
In the Brans-Dicke case of scalar-tensor theory ($\omega$=constant, \mbox{$V(\phi)=0$}) eq.(\ref{xiST}) simplifies considerably:
\begin{align}
\label{xiBD1}
\dx_{BD} (a)&=\left[\phi+\frac{\dot\phi}{\Hu}-X\right]\times\left[\phi+\frac{1}{2}\left(\frac{\dot\phi}{\Hu}-X\right)\right]^{-1}-1\\
\mathrm{where}\quad X&=\frac{\dot\phi}{\Hu}\frac{\left(\frac{\omega_{BD}\dot\phi}{\Hu \phi}-3\right)}{\left(\frac{\omega_{BD}\dot\phi}{\Hu \phi}-1\right)}\label{xiBD2}
\end{align}
Applying the condition $\omega_{BD}\gg 1$ (note that GR is recovered in the limit $\omega_{BD}\rightarrow\infty$) we have approximately:
\begin{align}
\delta\xi_{BD}(a)\sim\frac{1}{\omega_{BD}}
\end{align}
The expression for $\beta_{BD}$ can be trivially obtained by substituting the conditions $\omega=\omega_{BD}$, \mbox{$V(\phi)=0$} into eq.(\ref{betaST}).
\vspace{0.5cm}

\noindent\textit{$\mathbf{f\left(R\right)}$ \textbf{Gravity}}\\
We define the $f\left(R\right)$ action such that the GR limit is given by $f\left(R\right)=R$. Then:
\begin{align}
\dx_{f_R}(a)&=\frac{4}{3}\left(\frac{1}{f_R}-1\right)\\
\beta_{f_R}(a)&=\frac{2(\Hu^2-\dot\Hu)(1-f_R)+\ddot{f}_R-2\Hu\dot{f}_R}{3\Hu^2(1-f_R)-3\Hu\dot{f}_R+\frac{1}{2}a^2(Rf_R-f\left(R\right))}\label{betafR}
\end{align}
where $f_R=d\,f\left(R\right)/dR$.%
\vspace{0.5cm}

\noindent\textit{\textbf{Einstein-Aether Theory}}
\begin{align}
\dx_{AE}(a)&=-\frac{\left[\alpha-(c_1+c_3)\frac{\dot\Hu}{\Hu^2}+c_1\left(\frac{\dot\Hu}{2\Hu^2}-1\right)\right]}{\alpha-1+c_1\left(\frac{\dot\Hu}{2\Hu^2}-1\right)} \\
\beta_{AE}(a)&\approx\frac{\alpha}{2}\left(\frac{1}{(1-\Omega_M^{(0)})}-1\right)\label{betaAE}
\end{align}
where $c_i$ are parameters of the theory, \mbox{$\alpha=c_1+3c_2+c_3$}, and we have assumed $\alpha \ll 1$. This last condition is necessary to prevent extreme modifications to the effective gravitational constant that are already ruled out by present data. \vspace{0.5cm}

\noindent\textit{\textbf{DGP}}\\
We consider here the `normal' branch of DGP, since the self-accelerating branch suffers from ghostly pathologies \cite{Koyama2005,Gorbunov2006}, and is essentially ruled out by present data \cite{Fang:2008kc}.
\begin{align}
\dx_{DGP}(a)&=\frac{1}{3}\left[1+\frac{2}{3}\Hu \tilde{r}_c\left(2+\frac{\dot\Hu}{\Hu^2}\right)\right]^{-1}\\
\beta_{DGP}(a)&=\frac{1}{3\Hu\tilde{r}_c}\frac{1}{(1-\Omega_M^{(0)})}\left(\frac{\dot\Hu}{\Hu^2}-1\right)\label{betaDGP}
\end{align}
$\tilde{r}_c=r_c/a$ is the comoving crossover scale.
\vspace{0.5cm}

\noindent\textit{\textbf{Horndeski's Theory}}\\
Even in the simplified quasistatic limit, the relevant expressions for Horndeski's most general second-order scalar-tensor theory are non-trivial. Therefore we have chosen to relegate them to Appendix \ref{app:HD}, borrowing heavily from the results of \cite{Gleyzes2013,Bloomfield_HD}. 


\section{Conclusions}
\label{section:discussion}

We need to ready our tools for extracting the maximum amount of information from the next generation of large galaxy surveys. In this paper we have presented one such tool: a powerfully general and efficient method for calculating the density-weighted growth rate, $f\sigma_8(z)$, in modified gravitational scenarios. Our formalism bypasses the need for lengthy theory-specific calculations or multiple theory-specific growth rate codes.

Working at the level of linear perturbation theory, we have found that the response of the growth rate to departures from GR can be written as a Green's function-like integral over two contributions, a source term and a kernel. The source term depends on the deviations from GR under consideration; it encodes how modifications to the clustering of matter and the expansion rate both affect $f\sigma_8(z)$. The kernel is the same in all situations: it depends only on the properties of $\Lambda$CDM+GR, and acts as a weighting factor. It controls the extent to which the growth rate at a given redshift is affected by earlier non-GR behaviour.

As a result of expressing our calculation this way, we are able to clearly identify the degeneracy between the conventional source of modified gravity effects (the modified Newton-Poisson equation and the `slip relation', eqs.(\ref{Poisson}) and (\ref{slip})) and changes to the  background expansion. While measurements at different redshifts can help to mildly break this degeneracy, it is clear from our results that, contrary to what is usually claimed, measurements of the growth rate are simply not enough to distinguish modified gravity theories from models of dark energy. Geometric measures will play a crucial role in breaking this degeneracy.

From a practical point of view, our formalism has several \textit{modus operandi}. The conventional approach is to constrain the Lagrangian parameters of a fully-specified gravity theory. Using the formulae presented in this paper removes the need to calculate (both analytically and numerically) the perturbation theory for every gravity model of interest. Alternatively, one can turn the usual approach around, instead using the data to ascertain the departures from GR that are still allowable. This information can then be used to guide the development of new theories.

However, remaining agnostic about gravity when analysing the data incurs different penalties. It seems that the best option is to find a compromise between the generality of a parameterization and the usefulness of the constraints obtained. This kind of balancing act occurs frequently in model-independent analyses (for another  example see \cite{Tarrant}).

There remain a number of systematic effects to be mastered before growth rates can be used to make decisive statements about gravity theories; our exclusively linear formalism does not provide insight into these. It has been argued that the statistical scatter between current measurements of the growth rate is somewhat smaller than one might expect from the error bars cited \cite{Macaulay2011}, pointing to the need for a clear and accurate understanding of the systematic effects at play. Likewise, our linear formalism does not capture any novel nonlinear phenomena that might be present in an underlying theory, such as screening mechanisms. 

A logical extension of the work presented here would be to determine whether similar Green's function-like expressions can be derived for other relevant quantities, such as weak lensing shear and cross-correlations of the Integrated Sachs-Wolfe effect. That is, can we develop a simple plug-and-play toolbox for generating modified gravity observables? Such an item would be invaluable for observers working with fresh data, allowing them to make rapid analyses of gravity theories without being overburdened by model specifics. 

\section*{Acknowledgements}
We are grateful for helpful discussions with Phil Bull, Erminia Calabrese, Joanna Dunkley, Andrew Jaffe, Edward Macaulay and Andrew Pontzen. T. B. is supported by All Souls College, Oxford. P. G. F. acknowledges support from the Leverhulme Trust, STFC, the Beecroft Institute for Particle Astrophysics and Cosmology, and the Oxford Martin School. C. S. is supported by a Royal Society University Research Fellowship.

\appendix

\section{Perturbation of $\Omega_M(x)$}
\label{app:u_deriv}
\noindent Here we derive the relation eq.(\ref{delta_om}). 

First observe that the general fluid evolution equation:
\begin{align}
\rho^\prime&=-3\rho\,(1+\omega(x))
\end{align}
has the following solution, where we write \mbox{$\omega(x)=-1+\beta(x)$}: 
\begin{align}
\label{bjd}
\rho(x)&=\rho(0)\,e^{-3\int_0^x\beta(x^\prime)\,dx^\prime}
\end{align}

Now consider two universes. The first is a perfect $\Lambda$CDM model. In the second universe, the non-matter sector is not a true a cosmological constant; there is a modification to gravity that can be recast in the form of a perfect fluid with an evolving equation of state \cite{KunzSapone2006,Hu:2007fw}. Recall that nearly any gravity theory can be written in this form, even if the expression for the effective equation of state is extremely complex. Observational viability restricts that the equation of state can only differ from $-1$ by a small amount, ie. $\beta$ in eq.(\ref{bjd}) must be small for all $x$.

In the first universe we have:
\begin{align}
\label{appom1}
\Omega_M(x)&=\frac{\rho_M(x)}{\rho_M(x)+\rho_{\Lambda 0}}=\frac{1}{1+R\,e^{3x}} 
\end{align}
where we have used $\rho_M(x)=\rho_{M0}e^{-3x}$ and defined \mbox{$R=\rho_{\Lambda 0}/\rho_{M0}$}. In the second universe, denoting the energy density of the effective fluid by $\tilde{\rho}_X (x)$ and analogously defining \mbox{$\tilde{R}=\tilde{\rho}_{X0}/\tilde{\rho}_{M0}$}:
\begin{align}
\label{appom2}
\tilde{\Omega}^{(0)}_M(x)&=\frac{\tilde{\rho}_M(x)}{\tilde{\rho}_M(x)+\tilde{\rho}_X(x)}\nonumber\\
&=\frac{1}{1+\tilde{R}\,\mathrm{exp}\left[3x-3\int_0^x\beta(x^\prime)dx^\prime \right]}
\end{align}
where we have used eq.(\ref{bjd}).

We are interested in the small perturbation to the matter fraction $\Omega_M(x)$ that results from perturbing about a $\Lambda$CDM universe. This is given by the difference between eqs.(\ref{appom1}) and (\ref{appom2}). Since $\beta$ is small at all times we can expand:
\begin{align}
 e^{-3\int_0^x\beta (x^\prime) dx^\prime}&\approx 1-3\int_0^x \beta (x^\prime) dx^\prime = 1-3\,u(x)
 \end{align}
 where the last equality defines $u(x)$. Furthermore, we argue that $R=\tilde{R}$, because the ratio of the non-matter energy density to the matter energy density is an experimentally-determined quantity. Whether we are living in the $\Lambda$CDM or non-$\Lambda$CDM universe, we would simply measure one value for this ratio and call it $R$. 
 
 Collecting expressions then, we have:
\begin{align}
\delta\Omega_M(x)&=\tilde{\Omega}_M(x)-\Omega_M^{(0)}(x)\nonumber\\
&=\frac{1}{1+Re^{3x}[1-3u(x)]}-\frac{1}{1+Re^{3x}} \nonumber\\
&\approx \frac{3 Re^{3x}\,u(x)}{[1+Re^{3x}]^2}\nonumber\\
&=3\,u(x)\Omega_M^{(0)}(x)\,\left(1-\Omega_M^{(0)}(x)\right)
\end{align}
where the last step uses eq.(\ref{appom1}) to eliminate $Re^{3x}$ in favour of $\Omega_M^{(0)}$. This is the result stated in \textsection\ref{subsection:response_f}.

\section{Deviation Source Term for Horndeski's Theory}
 \label{app:HD}
 Given the complexity of the Horndeski Lagrangian \cite{Horndeski, Deffayetetal_2011, deFelice}, it is beyond the scope of this paper to carry out the intricate reduction to the quasistatic limit. Fortunately, this calculation has recently been presented in \cite{Gleyzes2013,Bloomfield_HD}, from which we adopt results. The calculation first appeared in \cite{Gleyzes2013} (see \cite{Gubitosi2013} for very closely-related work). However, the notation of \cite{Bloomfield_HD} is closer to that used in this paper, so we will use this as our source.
 
 In the expressions below $\bar{M}_i$ are parameters and $\Omega(\eta),\,\Lambda(\eta)$ and $c(\eta)$ are functions of conformal time that appear in the Lagrangian of the Effective Field Theory of Dark Energy -- we refer to \cite{Bloomfield2012, Bloomfield_HD} for precise definitions. $M_P$ is the Planck mass. The notational equivalence between metric potentials used in this paper and \cite{Bloomfield_HD} is $\Phi\equiv\psi$, $\Psi\equiv\phi$. We have also converted the expressions of \cite{Bloomfield_HD} from physical time to conformal time.
 \begin{widetext}
 \begin{align}
 \beta_{HD}(a)=&\frac{2(\Hu^2-\dot\Hu)(1-\Omega)+2\,M_P^{-2} a^2\,c+\ddot\Omega-2\Hu\dot\Omega}
 {3\Hu^2(2-\Omega_{M0}^{(0)}-\Omega)+M_P^{-2}a^2\left[2c-\Lambda\right]-3\Hu\dot\Omega}
 \label{betaHD}\\
 \dx_{HD}(a)=& \frac{B_\pi C_\Phi-B_\Phi C_\pi-B_\Phi C_{\pi2}\frac{a^2}{k^2}}
{A_\Phi\left(B_\Psi C_\pi+B_\Psi C_{\pi2}\frac{a^2}{k^2}-B_\pi C_\Psi\right)
+A_\pi\left(B_\Phi c_\Psi-B_\Psi C_\Phi\right)}
\end{align}
where:
\begin{align}
 A_\Phi&=2(M_P^2\Omega+\bar{M}_2^2) &
 A_\pi&=-(M_P^2\frac{\dot\Omega}{a}+\bar{M}_1^3)\\
 B_\Phi&=-1 &
 B_{\Psi}&=1+\frac{\bar{M}_2^2}{\Omega M_P^2}\\
 B_\pi&=\frac{\dot\Omega}{a\,\Omega}+\frac{\bar{M}_2^2}{\Omega M_P^2\,a}\left(\Hu+2\frac{\dot{\bar{M}}_2}{\bar{M}_2}\right) &
 C_\Phi&=M_P^2\frac{\dot\Omega}{a}+\frac{\bar{M}_2}{a}\left(\Hu+2\frac{\dot{\bar{M}}_2}{\bar{M}_2}\right)\\
C_\Psi&=-\frac{M_P^2}{2}\frac{\dot\Omega}{a}-\frac{\bar{M}_1^3}{2} &
C_\pi&=c-\frac{\bar{M}_1^3}{2a}\left(\Hu+3\frac{\dot{\bar{M}}_1}{\bar{M}_1}\right)+\frac{\bar{M}_2^2}{a^2}\left(2\Hu^2-\dot{\Hu}+2\Hu\frac{\dot{\bar{M}}_2}{\bar{M}_2}\right)
\end{align}
\begin{align}
C_{\pi 2}&=\frac{M_P^2}{4a^2}\dot\Omega \dot{R}^0-\frac{3 c}{a^2} \left(\dot{\Hu}-\Hu^2\right)+\frac{3\bar{M}_1^3}{2 a^3}\left(3\frac{\dot{\bar{M}}_1}{M_1}(\dot{\Hu}-\Hu^2)+\ddot{\Hu}-\Hu\dot{\Hu}-\Hu^3\right)+3\frac{\bar{M}_2^2}{a^4}\left(\dot{\Hu}-\Hu^2\right)^2
 \end{align}
 \end{widetext}



\bibliographystyle{apsrev}

\end{document}